\documentclass[preprint,showpacs,preprintnumbers,amsmath,amssymb]{revtex4}

\usepackage{graphicx}
\usepackage{dcolumn}
\usepackage{bm}
\usepackage{citesupernumber}

\begin{document}

\preprint{APS/123-QED}

\title{Directionality is an inherent property of biochemical networks}

\author{Feng Yang}
\email{fyang@mcw.edu}
\author{Feng Qi}
\author{Daniel A. Beard}
\email{dbeard@mcw.edu}
\affiliation{Biotechnology and
Bioengineering Center, Department of Physiology, Medical College
of Wisconsin, Milwaukee, Wisconsin 53226, USA}

\date{\today}

\begin{abstract}
Thermodynamic constraints on reactions directions are inherent in
the structure of a given biochemical network. However, concrete
procedures for determining feasible reaction directions for
large-scale metabolic networks are not well established. This work
introduces a systematic approach to compute reaction directions,
which are constrained by mass balance and thermodynamics, for
genome-scale networks. In addition, it is shown that the nonconvex
solution space constrained by physicochemical constraints can be
approximated by a set of linearized subspaces in which mass and
thermodynamic balance are guaranteed.  The developed methodology
can be used to {\it ab initio} predict reaction directions of
genome-scale networks based solely on the network stoichoimetry.
\end{abstract}


\maketitle


Constraint-based approaches to analyzing biochemical networks,
such as flux balance analysis (FBA) and energy balance analysis
(EBA), have found widespread applications in system analysis of
metabolic networks \cite{Joyce06}. A successful set of procedures
has been established for determining metabolic reactions which are
present in a genome-scale system \cite{Forster03, Francke05}.
However, the procedure for determining feasible reaction
directions is less concrete and arbitrary to some degree
\cite{Kummel06,Price06}. Yet the information on reaction
directions is crucial to ensure that predicted reaction fluxes are
physically feasible and obtained results are physiologically
reasonable \cite{Beard02, Beard04a}.

Allowable reaction directions in a biochemical network are
constrained by mass-conservation constraints and thermodynamic
constraints, both arising from the stoichiometry of a given
network \cite{Lee06, Beard02,Henry06,Price06}. While the linear
mass-conservation constraint is rigorously applied in many
applications, the nonlinear thermodynamic constraint is not
typically considered due to its implicit nonlinearity and
NP-completeness of the associated problems. Consequently, the flux
distribution determined solely by the principle of
mass-conservation, may not be physically feasible.

The nonlinear thermodynamic constraint is based on the second law of
thermodynamics which requires that each internal reaction with
non-zero flux must dissipate energy \cite {Beard04a}. For a set of
reaction directions to be thermodynamically feasible (T-feasible),
there must exist a thermodynamic driving force for the reaction
directions on a given network. A robust algorithm to implement this
constraint, which is based on an NP-complete computation of entire
elementary modes for a network, has been previously developed
\cite{Yang05}. However, an appropriate method for genome-scale
networks has not been developed until now. Here, we introduce a new
methodology for systematically determining feasible reaction
directions for large-scale networks and apply it to a biochemical
network model of the mammalian cardiomyocyte \cite{Vo06}.



To illustrate mass-balanced and thermodynamically-defined
determination of feasible reaction directions, an example network
illustrated in Fig. 1 is used.
\begin{figure}
\scalebox{0.65}[0.65]{\includegraphics{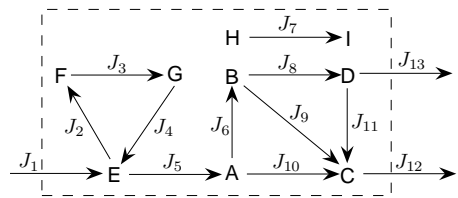}}
\caption{\label{fig:toy_nework} A example network composed of 13
reactions ($J_{1-13}$) and 9 species (A through I). $J_1$,
$J_{12}$ and $J_{13}$ are transport (boundary) fluxes that
transport mass into and out of the system; $J_2$ through $J_{11}$
are internal fluxes, as shown in the dashed box.}
\end{figure}
In this network, transport fluxes $J_1$, $J_{12}$, and $J_{13}$
are assumed to have net flux in the direction indicated by the
arrows in the figure (defined as positive). In general, we assume
that some subset of reaction fluxes (typically transport fluxes)
have associated with them prior assigned directions. A set of
these predefined constraints is denoted as $\Psi$. For example of
Fig. 1, the constraints on boundary fluxes can be expressed
$\Psi$: sign($J_1$, $J_{12}$, $J_{13}$) = \{+, +, +\}.


 {\bf Flux directions constrained by mass-balance only}
For a flux vector $\vec J$ to be feasible, steady-state mass
balance requires
\begin{equation}
   \sum_j S_{ij} \cdot J_j = 0, \text{ and } \vec J \text{ obeys } \Psi  \\
\end{equation}
where $S_{ij}$ is the stoichiometric coefficient of metabolite $i$
and reaction $j$. {\it Mass-irreversible} ($MI$) fluxes are those
fluxes, the signs of which are specified by considering the
mass-balance and boundary constraints of Eq. 1 alone. For example
of Fig. 1, $J_5$ is mass-irreversible, because there exists no
feasible flux vector $\vec J$ that satisfies Eq. 1 for which
$J_5<0$. In fact, here we say flux $J_5$ is {\it strictly
mass-irreversible} ($sMI$) because $J_5>0$ and the equality
$J_5=0$ is not feasible for finite boundary flux values. Such
fluxes represent the backbone of the network; without them, the
network would not function with any nonzero fluxes. In addition,
$J_7$ is characterized as {\it mass-infeasible} ($MN$) because no
flux is possible for this network given the boundary conditions
and the network stoichiometry. The remaining internal reactions in
Fig. 1 ($J_{2-4},J_6,$ and $J_{8-11}$) are denoted as {\it
mass-reversible ($MR$)}.

{\bf Flux directions constrained by thermodynamic
constraints} The nonlinear thermodynamic constraint requires that
there exists a driving force $\vec \mu$ for a given flux vector
$\vec J$ such that
\begin{eqnarray}
&&\Delta \mu {_j} = \sum_i \mu {_i} \cdot S_{ij}, \hspace{1 mm} J_j
\cdot \Delta \mu {_j} \leq 0,  \nonumber\\
&&\text{and} \hspace{1 mm} J_j = 0 \hspace{1 mm}  \hspace{1 mm}
\text{if and only if} \hspace{1 mm} \Delta \mu {_j} = 0, \hspace{1
mm}
\end{eqnarray}
where $i$ and $j$ are the indexes for a particular metabolite and
internal reaction, respectively. {\it Thermodynamically
irreversible} ($TI$) fluxes are those identified as mass-reversible
by Eq. 1 but found to be irreversible when T-constraints (Eq. 2) is
imposed. In Fig. 1, $J_6$, $J_8$ and $J_{10}$ are $TI$ fluxes, with
their feasible net fluxes in the positive directions illustrated in
the figure. There exists no T-feasible flux patterns for which these
reactions operate with net fluxes in the negative direction. In
addition, reactions $J_2$, $J_3$ and $J_4$ are characterized as {\it
T-infeasible} ($TN$) because no net flux is possible in any of flux
directions; reactions $J_9$ and $J_{11}$ can proceed in both
directions while still satisfying mass-balance and T-constraints,
and thus they are denoted as {\it T-reversible} ($TR$) reactions. An
algorithm for determining $TI$, $TN$, and $TR$ fluxes in a given
network is presented in the Supplemental Material.

In characterizing the impacts of applying T-constraints, we observe
that a mass-reversible ($MR$) reaction is potentially {\it
reducible} by T-constraints, either reduced as T-irreversible ($TI$)
or reduced as T-infeasible ($TN$) . In addition, a mass-irreversible
($MI$) reaction is also {\it reducible} if it is later characterized
as T-infeasible, though it may not be necessarily true.
Furthermore, the constraints of Eq. 2 can be equivalently
represented by the orthogonality of a flux vector $\vec J$ against
the sign patterns of the exhaustive set of infeasible reaction
cycles \cite{Beard04a}. Here, a {\it reaction cycle} is a set of
reactions, the summation of which results in cancellation of all
participating metabolites. There are 4 reaction cycles imbedded in
the example network (Fig. 1), for example, three fluxes of $J_2$,
$J_3$, and $J_4$ make up a cycle:
 $J_2: E \rightleftharpoons F; 
 J_3: F \rightleftharpoons G; 
 J_4: G \rightleftharpoons E, $ 
 and thus $J_2+J_3+J_4: \varnothing = \varnothing$.
{\bf Flux directions that are sufficient for T-feasibility} The
overall goal of {\it ab initio} prediction is to generate a set of
irreversibility constraints that guarantees the T-feasibility.
However, the constraints defined so far ($MI$ and $TI$) are merely
necessary, but not sufficient to ensure the T-feasibility in general
\cite {Yang05}. Consequently, the solution space defined by $MI$ and
$TI$ is not linearly constrained and still a nonconvex space. For
instance, to generate a linearly constrained, T-feasible solution
space for the example network, one of the following additional
linear constraints has to be included: I) If $J_9>0$, then $J_{11}$
is feasible in any directions; II) If $J_{11}<0$, then $J_9$ is
feasible in any directions; III)  If $J_9<0$ and $J_{11}< 0$; IV) If
$J_9>0$ and $J_{11}>0$, $etc$.  The resulting convex spaces
($_lEBA$), as well as traditional flux balance space ($FBA$) and
nonconvex T-feasible space($_fEBA$), are illustrated in Fig. 2.
\begin{figure}
\scalebox{0.75}[0.75]{\includegraphics{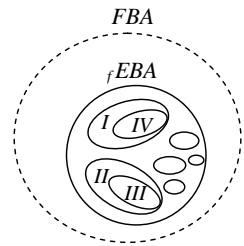}}
\caption{\label{fig:FBA_EBA} Illustration of the solution space
constrained by mass-balance only ($FBA$), the space constrained by
additional nonlinear T-constraints ($_fEBA$), and the linearlized
T-feasible spaces ($_lEBA$, such as the solution spaces of I, II,
III, IV, $etc$). The space defined by traditional $FBA$ may not be
physically feasible; the space of $_fEBA$, in which both
mass-balance and energy-balance are guaranteed, represents a better
approximation of network behaviors than using $FBA$ only.}
\end{figure}

In general, in addition to those necessary conditions we need to
determine a set of extra irreversibility constraints ($EI$) to
guarantee the T-feasibility. The resulting linearized solution
spaces ($_l{EBA}$) defined by $MI$, $TI$, and $EI$ are subsets of
the nonconvex T-feasible space ($_f{EBA}$). However, a complete
enumeration of these subspaces is not practicable (maximally
{$21^{n}$}, where $n$ is the number of T-reversible fluxes in a
given network), especially for large-scale networks. To approximate
the nonconvex $_fEBA$ space, however, it is possible to enumerate a
limited set of subspaces ($_l{EBA}$) in which the number of
reversible reactions is maximized. The subspaces obtained represent
the largest subsets of $_fEBA$, such as the spaces of I and II in
Fig. 2. Therefore, the network behaviors could be maximally captured
by studying these linearized subspaces. For those subspaces that are
heavily constrained, for instance, the spaces of III, IV and many
others, it is assumed that they represent the non-typical network
behaviors, or the time staying in these states is highly limited. In
practice, determining which set of $_l{EBA}$ subspaces for
particular studies may require further constraints or evidences from
experimental observations.



In summary, for a given network with associated prior reaction
directions, to generate a convex, T-feasible solution space,
reaction directions can be categorized as \{MN, sMI, MI, TN, TI, TR,
and EI\}. A diagram which shows their relationship are illustrated
in Fig. 3.
\begin{figure}
\scalebox{0.50}[0.50]{\includegraphics{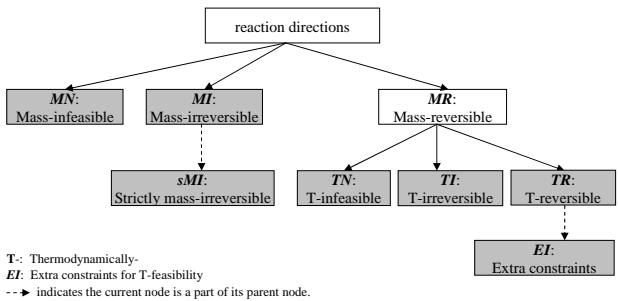}}
\caption{\label{fig:FBA_EBA} Illustration of the categories of
reaction directions (shaded in gray) when both mass-conservation
and T-constraints are imposed. }
\end{figure}
In particular, when all reaction directions other than those of
predefined boundary fluxes are assumed reversible, a few theorems
have been observed in the above analysis.

%



{\bf Theorem 1 Mass-irreversible reactions are not reducible by
thermodynamic constraints.} Suppose a $MI$ reaction that connects
species A to B, $A \rightarrow B$, is reduced as T-infeasible by a
reaction loop which connects species B to A along $L$, $B
\rightarrow ^L \rightarrow A$. As a result, there must be a pathway
that connects A to B along a reversed loop of $L$, $A \rightarrow
^{-L} \rightarrow B$, by the definition of mass balance. The
existence of such a reversed loop thus disables the reducibility of
the loop $L$, and there is no cycle existing between species A and
B. This theorem implies a mass-irreversible reaction does not
participate in any reaction cycle; if involved, then it must be
characterized as mass-reversible by Eq. 1. In other words, $MI \cap
TI = \varnothing $, where $MI$ and $TI$ represent the sets of
mass-irreversible and T-irreversible reactions, respectively.


{\bf Theorem 2 Mass-reversible reactions can be reduced as
T-irreversible or T-infeasible.} Supposing a reversible flux $J_i$,
$A \leftrightarrows B$, involved in a loop $L$, then all reactions
in this loop $L$ must be reversible because of the connectivity
between A and B along $L$. To block this reaction cycle, any
reversible reactions can be chosen and reduced as irreversible.
Therefore, the loop constraint is active with respect to the
reversible flux $J_i$, which can be reduced as T-infeasible (such as
$J_{2-4}$), or reduced as T-irreversible (such as $J_{6,8,10}$ in
the example network).

{\bf Theorem 3  A reaction cycle in which all reactions are
irreversible is impossible.} Suppose there is a reaction cycle
which contains an irreversible reaction $A \rightarrow B$, and a
set of irreversible reactions along loop $L$, then the reactions
along loop $L$ must be reversible by mass-balance. Therefore, a
fully directed cycle is impossible and any cycles can always be
eliminated by assigning directions to some T-reversible reactions
in the network.

The developed methodology has been applied for the metabolic
network of mammalian cardiomyocyte \cite{Vo06} which accounts for
$240$ metabolites and $257$ reactions. While the stoichiometric
structure of this model has been adopted, the flux directions
previously assigned are typically not considered. Instead, we
define a network in which the directions of all boundary fluxes
and two internal fluxes (ATP hydrolysis and a transport flux of
$12DGRt2$), are assigned based on the published model \cite{Vo06},
but all other reactions are assumed to be reversible in default.
In this network model, three transport fluxes are constrained to
operate in the direction of net uptake of substrates (glucose,
oleate and oxygen) and 28 transport fluxes (including citrate,
acetate and 26 additional output fluxes) are constrained to
operate in the direction of net production of reactants.  A total
of 33 irreversibility constraints are specified and the directions
of the remaining 217 internal reactions are computed.

Given finite uptakes of glucose, oleate and oxygen, mass-balance
analysis identifies 18 $sMI$ internal fluxes (see red and blue
fluxes in Fig. 4),
\begin{figure}
\scalebox{0.35}[0.35]{\includegraphics{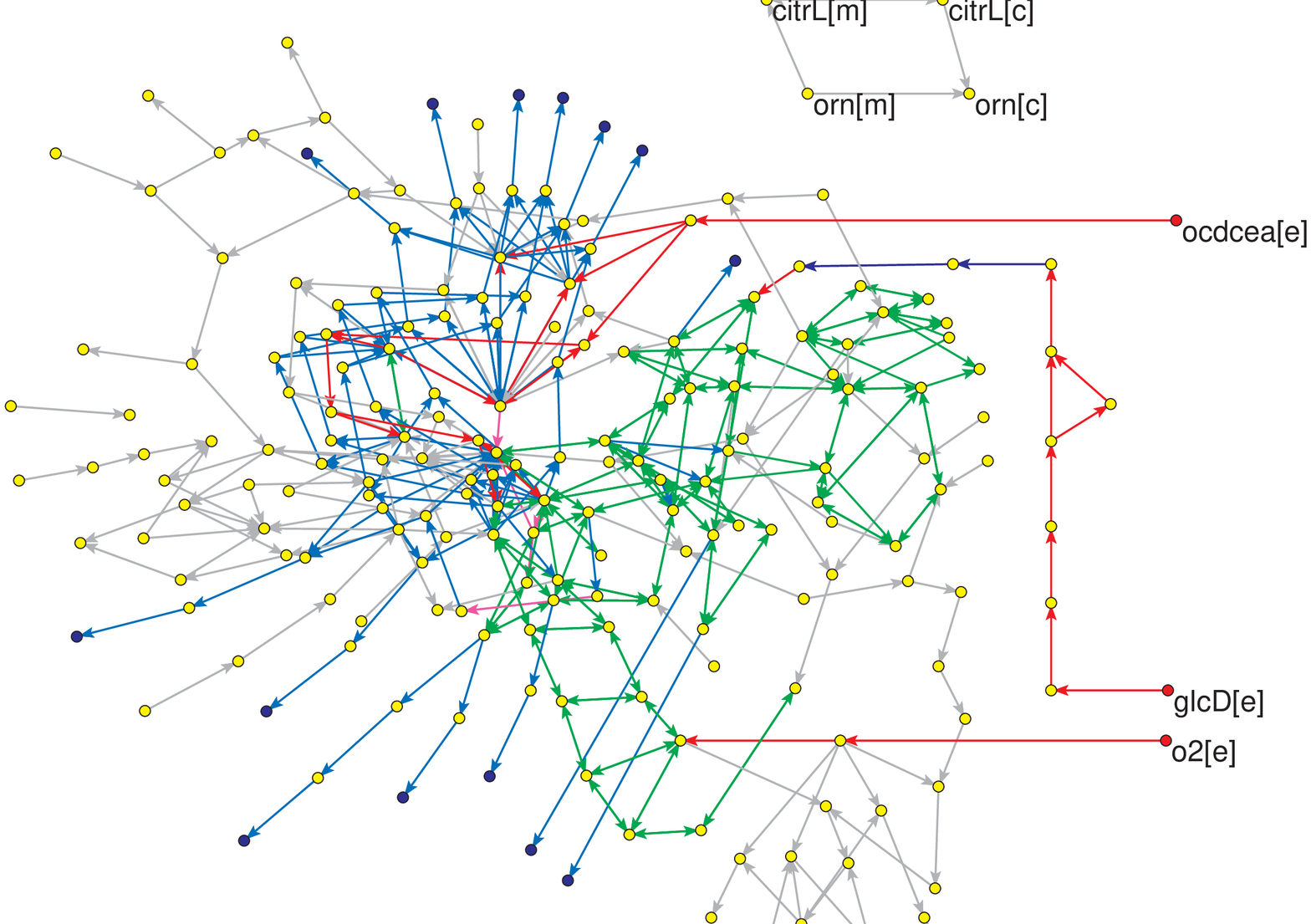}}
\caption{\label{fig:T_network} Illustration of a thermodynamically
feasible network of the cardiomyocyte network. The nodes and edges
represent metabolites and reactions, respectively. In particular,
the red, blue and yellow nodes represent the sources, the sinks and
the intermediate metabolites of the network, respectively. Red edges
and blue edges represent the strictly mass-irreversible reactions,
and violet red edges and navy blue edges represent mass-irreversible
reactions. In particular, reactions in red or violet red proceed in
the same direction as the original definition of the network
\cite{Vo06}; reactions in blue and navy blue proceed in a reversed
direction. An infeasible reaction cycle is shown at the right top
corner of Fig. 4. For simplicity, the currency of the biochemical
networks, such as ATP, ADP, Pi, NADH, NAD, H$^+$, H$_2$O and CO$_2$,
have not been shown.}
\end{figure}
which represent a set of reactions essential for this network. It
also shows that only 14 products (of 32 desired metabolites) could
be possibly produced from the sources (i.e. glucose, oleate, and
O$_2$), with the aids of three reversible transport fluxes of
CO$_2$, H$^+$, and H$_2$O. In addition, 70 mass-infeasible fluxes,
73 mass-irreversible fluxes, and 74 mass-reversible fluxes are
found. Furthermore, the results of intensively sampled T-feasible
networks (559,518,687 samples) further indicate that among 74
mass-reversible fluxes, three fluxes ($CITRtm$, $ORNt3m$,
$ORNt4m$) are T-infeasible, because these three reactions
\begin{eqnarray}
 && CITRtm:  citrL[m] \rightleftharpoons citrL[c], \nonumber \\
 && ORNt3m:    h[c] + orn[m] \rightleftharpoons h[m] + orn[c],  \nonumber \\
 && ORNt4m: citrL[c] + h[c] + orn[m] \rightleftharpoons
 citrL[m] + h[m] \nonumber\\
&&  + orn[c]  \\
\nonumber
\end{eqnarray} develop a reaction cycle as $CITRtm + ORNt4m -
RNt3m = \varnothing$. It is also found that $h[c]$ is the only
metabolite that connects this reaction cycle to the remaining part
of the network, which reminds of the reaction cycle $J_2
\rightarrow J_3 \rightarrow J_4 \rightarrow J_2$ in the example
network. Thus, neither direction is feasible for all these three
reactions.

The solution space defined by $MI \cup TI$ needs to be further
refined since 1099 cycles are still embedded in this network.
However, by specifying only 14 reversible fluxes (out of 71
T-reversible fluxes) be irreversible, it is possible to eliminate
all those cycles. A final T-feasible network for mammalian
cardiomyocyte has been illustrated in Fig. 4, which consists of 112
infeasible reactions, 101 irreversible reactions and 44 reversible
reactions. However, such a linearly constrained subspace is not
unique, as indicated in the example network. To further explore the
nonlinear T-feasible space, a large number of subspaces (see Fig. 5)
\begin{figure}
\scalebox{0.35}[0.35]{\includegraphics{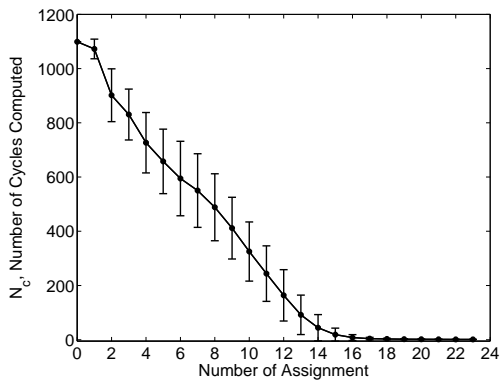}}
\caption{\label{fig:UI_fig} Demonstration of reductions in the
number of reaction cycles with extra irreversibility constraints
($EI$). It represents the results of $10^5$ independent simulations
for generating T-feasible solution spaces. Bars represent the
standard deviation of the number of cycles that are eliminated at
each particular assignment. }
\end{figure}
have been generated by assigning directions to those T-reversible
fluxes. Importantly, each irreversibility assignment is generated in
such a way that the maximum number of reaction cycles can be
eliminated. Therefore, the generated set of linear constraints
defines a feasible solution space which requires a minimal number of
irreversibility constraints, and thus represents the largest linear
subspace in Fig. 2.

Surprisingly, the computed T-feasible irreversibility constraints
show significant difference when compared to the originally
assigned directions, by which the space defined is not
thermodynamically feasible (7 reaction cycles are included). To
make it physically feasible, three mass-reversible fluxes
($CITRtm$, $ORNt3m$, $ORNt4m$) have to be eliminated, and at least
one additional direction assignment, either specifying $AKGMALtm$
(previously assigned as reversible) be positive or specifying
$ASPTAm$ be negative, has to be incorporated into the original
definition of the network. Although these additional constraints
do not invoke significant impacts on the overall results from
\cite{Vo06}, they do change the flux distribution with the given
boundary fluxes. While the revised network represents a special
subset of the T-feasible solution space, it is significantly
different from the network which shows the maximum level of
flexibility (i.e. containing the maximum number of reversible
reactions). Table 1 shows the difference between an example of the
largest T-feasible subspaces, and the original definition of the
network (PI) and its mass-balanced network (PI+FBA) and
thermodynamically balanced network (PI+FBA+EBA).
\begin{table}
\caption{\label{tab:table1}Comparison of a T-feasible network with
high flexibility to the original network as well as its
corresponding mass-balanced and thermodynamically balanced
networks. }
\begin{ruledtabular}
\begin{tabular}{cccccccc}
 &PI  &PI+FBA    &PI+FBA+EBA  &EBA\footnotemark[1]  \\
 \hline
 Infeasible fluxes& 0 & 149 & 149 & 112 \\

 Positive\footnotemark[2] fluxes& 156 & 60 & 61 & 42 \\

 Negative\footnotemark[2] fluxes& 3 & 21 & 21 & 59 \\

 Reversible fluxes& 98 & 27 & 26 & 44 \\
\end{tabular}
\end{ruledtabular}
 \footnotetext[1]{A T-feasible network corresponds to the network shown in Fig. 4.}
 \footnotetext[2]{The positive reaction direction is defined based on the left-to-right reaction
 direction in the original model definitions \cite{Vo06}.}
\end{table}
The significant difference between the {\it ab initio} prediction
and the previously assigned directions, can be explained by the
fact that mass-conservation and thermodynamic constraints are by
far not the only constraints that a given network must satisfy.
Those nonidentified may come from gene regulatory constraints,
topological constraints, environmental constraints
\cite{Palsson06}, and constraints that are accumulated from
evolution and natural selection. By systematically studying the
difference between the results from {\it ab initio} predictions
and experimental observations, we may find some interesting
knowledge ``gap", leading to more accurate predictions on the flux
distributions.

In summary, a biophysical constraint is applied to genome-scale
biochemical networks for computing flux directions and determining
a linearly constrained, mass- and thermodynamically-balanced
solution space. Given a reaction network and a set of prior
constraints on certain flux directions (such as known transport
flux directions) the algorithm determines a minimal set of
irreversibility constraints (reactions directions) that is
sufficient to guarantee T-feasibility. The developed methods could
be used as a tool to systematically design a biochemical network,
greatly facilitate the further flux balance analysis and energy
balance analysis, improve their predictions of the flux
distribution, and help capturing the physically meaningful
behavior of the network studied. Numerically, a linear solution
space will make related optimizations more reliable and efficient.

\begin{acknowledgments}
I wish to acknowledge the support of all authors, although Daniel
A. Beard does not like the overall organization of this
manuscript.
\end{acknowledgments}

\bibliographystyle{unsrt}
\bibliography{directionality}

\end{document}